\documentclass{JHEP3}
\newcommand{\beq}{\begin{equation}}
\newcommand{\eeq}{\end{equation}}

\renewcommand{\d}{\delta}
\renewcommand{\c}{\chi}
\usepackage{epsfig,multicol}

\title{Lattice Sigma Models with Exact Supersymmetry}

\author{Simon Catterall and Sofiane Ghadab\\
        Department of Physics, Syracuse University, Syracuse, NY 13244, USA\\
        E-mail: \email{smc@physics.syr.edu}\\
        E-mail: \email{sghadab@physics.syr.edu}}

\preprint{SU-4252-789}  

\abstract{We show how to construct lattice sigma models in one, two
and four dimensions which exhibit
an exact fermionic symmetry. These models are discretized and {\it twisted}
versions of conventional supersymmetric sigma models with $N=2$
supersymmetry. The fermionic symmetry corresponds to a scalar BRST
charge built from the original supercharges. 
The lattice theories possess local actions and in many cases
admit a Wilson term to suppress doubles. In the two and four dimensional theories we show
that these lattice theories are invariant under additional 
discrete symmetries.
We argue that the presence of these exact symmetries
ensures that no fine tuning is required to achieve $N=2$ supersymmetry
in the continuum limit.
As a concrete example we show preliminary numerical results
from a simulation of the $O(3)$ supersymmetric sigma model in two
dimensions.
}


\begin{document} 
\section{Introduction}
Supersymmetric field theories exhibit many remarkable properties. Chief
among these are the cancellations which occur between boson and
fermion contributions in the perturbative calculation of physical
quantities. These cancellations eliminate many of the divergences
typical of quantum field theory and are at the heart of its use to
solve the gauge hierarchy problem \cite{wein}. Additionally, supersymmetric
versions of Yang Mills theory, while more tractable analytically than their
non-supersymmetric counterparts, exhibit many of the same features such as
confinement and chiral symmetry breaking \cite{prop}. Finally, supersymmetric $SU(N)$
gauge theory in the large $N$ limit has been proposed 
as a candidate for M-theory \cite{M}. 

Given then,
the theoretical and phenomenological interest in supersymmetric
theories, it is perhaps not surprising that a good deal of
effort has gone into attempts to study such theories on spacetime
lattices see, for example, \cite{old,recent} and the recent reviews
by Feo and Kaplan \cite{feo,kaplan}.  
However, until recently these efforts mostly met with only limited
success. The reasons for this are well known -- generic discretizations of supersymmetric field
theories break supersymmetry at the classical level leading to the
appearance of a 
plethora of {\it relevant} susy breaking counterterms in the effective
action. 
The couplings to all these terms must then be fine tuned as the lattice
spacing is reduced in order that the theory approach a supersymmetric
continuum limit. This problem is particularly acute in theories
with extended supersymmetry which contain scalar fields.
A notable exception to this scenario is $N=1$ super Yang Mills in four
dimensions in which it was shown that only one coupling need be tuned
to restore supersymmetry in the continuum limit \cite{ven}. Because of
this a good deal of numerical work has been devoted to this model
\cite{montway,feo}
.
In \cite{wz2,qm} we showed that in certain cases this fine tuning problem
could be evaded by constructing a lattice theory which retained
an element of supersymmetry. A similar philosophy has been followed
in recent papers by Kaplan et al. \cite{kap1,kap2,kap3}. The latter work
allows constructions of lattice super Yang Mills models by orbifolding
a large $N$ matrix model. In \cite{top} we showed that the models
treated in \cite{wz2,qm}, namely supersymmetric quantum mechanics and
the complex Wess-Zumino model, were actually examples of a wider
class of supersymmetric theory -- namely theories which could be
{\it twisted} to expose a scalar fermionic symmetry. 
This fermionic symmetry can be written in terms of a charge $Q$, formed
from combinations of the original supercharges, and 
obeying the BRST-like algebra $Q^2=0$. We will refer to it
as a {\it twisted supersymmetry}. Furthermore, the resulting
twisted actions generically
take the form $S=\delta \Lambda$ where $\delta$ is the variation generated
by $Q$ and $\Lambda$ is some function of
the fields. They resemble pure gauge fixing
terms. It is important to realize
that in flat space this
twisting operation can be considered as merely a relabeling of the
fields of the theory and hence does not alter its physical content.
   
This twisting procedure is very well known
in the literature, as the first stage in the
construction of {\it topological field theories} (see for example
the reviews \cite{rev} and \cite{rev2} and references
therein). Indeed Witten pioneered the technique when he twisted
$N=2$ super Yang Mills to arrive at the very first topological field theory
\cite{witten1}.  
The final transition from twisted supersymmetric theory to
true topological field theory necessitates restricting the
set of physical observables to only those invariant under the BRST symmetry. 
In the language of the parent supersymmetric field theory this
constitutes a projection to the ground state of the theory and is responsible
for ensuring that the topological field theory does not contain
local degrees of freedom.
In our constructions of lattice supersymmetric models we will {\it not} impose
this restriction and our actions are simply discrete versions of the
continuum twisted supersymmetric actions.

As we have remarked, the structure of these twisted actions suggests that
the associated topological field theories 
can be obtained by a gauge fixing procedure. 
This is
indeed the case and we will construct our lattice actions by
writing down an appropriate BRST symmetry and making a suitable
choice of gauge fixing condition.  
We will show that the resulting
lattice models enjoy a number of nice properties. They are local,
yield a $N=2$ supersymmetric model in the classical continuum limit and
most importantly maintain an exact (twisted) supersymmetry.  We
argue that the presence
of this exact symmetry should protect the theory from
the usual radiative corrections which plague lattice supersymmetric
models.
The construction is sufficiently general
that we can, with certain restrictions on the target space, formulate
the models in one, two or four dimensions. These restrictions on the
target space are nothing more than the usual restrictions required for
the theories to exhibit extended supersymmetry.

In the next section we describe the procedure used to construct these lattice
sigma models
focusing initially on the case of a one dimensional base space (yielding
a generalization of the quantum mechanics model considered earlier \cite{wz2}).
The following section shows how to generalize the model to two
dimensions leading to a lattice version of the usual supersymmetric
sigma models with extended supersymmetry. 
In this case the requirement that the BRST invariant model possess
an interpretation as a twisted form of a supersymmetric theory requires
that the target space be K\"{a}hler. 
We then discuss in some detail perhaps the simplest example
of such a model -- the $O(3)$ sigma model. Some preliminary numerical
results which support our reasoning are then shown. 
Finally, the natural generalization
to four dimensions is given which requires that the target manifold be
hyperK\"{a}hler. 

Most of the results derived in this paper are not
new but have existed
in the literature of topological field theory for a decade or more.
The crucial
observation we make here is 
that many of them can be taken over without significant
modification to the lattice. The purpose of this paper is to
explain how this comes about in as pedagogical a way as possible. Further
discussion of continuum
topological field theories and
their relationship to
supersymmetric theories can be found in the excellent review \cite{rev} and
references therein. 

\section{A BRST invariant model on a curved target space}

Consider a set of real bosonic fields $\phi^i(\sigma)$ corresponding to
coordinates on an $N$-dimensional target space equipped with 
Riemannian metric
$g_{ij}(\phi)$. The coordinates $\sigma$ parametrize a base space whose
dimension and other properties will be determined later. In addition
to $\phi^i$ let us introduce real {\it anticommuting} fields
$\psi^i(\sigma)$ and $\eta_i(\sigma)$ and a further bosonic (commuting) field
$B_i(\sigma)$. These fields will transform as
vectors in the target space. Let us further postulate the following set of transformations
parametrized by a single (global) grassmann parameter $\xi$
\begin{eqnarray}
\label{aa}
\delta\phi^i&=&\xi\psi^i \nonumber \\
\delta\psi^i&=&0 \nonumber \\
\delta\eta_i&=&\xi\left(B_i-\eta_j\Gamma^j_{ik}\psi^k\right) \nonumber \\
\delta
B_i&=&\xi\left(B_j\Gamma^j_{ik}\psi^k-\frac{1}{2}\eta_jR^j_{ilk}\psi^l\psi^k\right)
\label{BRST}
\end{eqnarray}
The quantities $\Gamma^i_{jk}$ and $R^i_{jkl}$ are the usual
Riemannian connection and curvature.
It is clear that $\delta^2\phi^i=\delta^2\psi^i=0$. It is straightforward,
(the details are given in appendix A) to prove the additional results
\begin{eqnarray}
\delta^2\eta_i&=&0\nonumber\\
\delta^2 B_i&=&0\\
\end{eqnarray}
Thus the symmetry is nilpotent. Indeed, as was shown in \cite{singer}, these
transformations can be regarded as a BRST symmetry which arises from
quantizing a purely bosonic model with trivial classical action $S_{\rm
cl}(\phi)=0$. This classical action is invariant under local
shifts in the field $\phi^i\to\phi^i+\epsilon^i$. BRST quantization
is then employed to define the quantum theory.
Thus, in this approach, the anticommuting fields $\psi^i$ and
$\eta_i$ are to be interpreted as a ghost and antighost field
respectively while the
field $B_i$ is a Lagrange multiplier field used to implement a 
(yet to be specified) gauge fixing
condition. Notice that none of these properties depend on the
nature of the base space.

Having chosen a field content and an associated BRST symmetry it now
remains to choose an action. Again, the gauge fixing perspective suggests
the following natural choice
\begin{equation}
\xi S=\alpha
\delta\left(\int_\sigma \eta_i\left(N^i\left(\phi\right)-\frac{1}{2}g^{ij}B_j\right)\right)
\end{equation}
The vector $N^i(\phi)$ corresponds to an arbitrary gauge fixing condition
on the bosonic field $\phi^i$. Notice that the exact nature
of the integral over the base space need not be specified at this point.
The coupling constant $\alpha$ corresponds in this
language to a gauge parameter. Expectation values
of observables which are invariant under the $\delta$-symmetry will
not depend on the value of $\alpha$. 
Notice also that {\it only} the nilpotent property of
the $\delta$ operator is needed to show the
the action is BRST invariant. 
Carrying out the variation leads to the following expression for
an invariant action
\begin{equation}
S= \alpha\int_\sigma
\left(B_iN^i-\frac{1}{2}g^{ij}B_iB_j-\eta_i \nabla_kN^i\psi^k+
\frac{1}{4}R_{jlmk}\eta^j\eta^l\psi^m\psi^k\right)
\end{equation}
where the symbol $\nabla$ indicates a target space covariant derivative.  
Integrating out the multiplier field $B$ yields the on shell action
\begin{equation}
S=\alpha\int_\sigma
\left(\frac{1}{2}g_{ij}N^iN^j-\eta_i \nabla_kN^i\psi^k+
\frac{1}{4}R_{jlmk}\eta^j\eta^l\psi^m\psi^k\right)
\end{equation}
This action is manifestly general coordinate invariant
with respect to the target space coordinates. It is also
invariant under the on shell fermionic transformations
\begin{eqnarray}
\delta\phi^i&=&\xi\psi^i\nonumber\\
\delta\psi^i&=&0\nonumber\\
\delta\eta_i&=&\xi\left(g_{ij}N^j-\eta_j\Gamma^j_{ik}\psi^k\right)
\end{eqnarray}
Up to this point we have left the choice of the target space vector
field arbitrary. In the case of a one dimensional base space
with coordinate $\sigma$ there is
a natural choice for $N^i\left(\phi\right)=\frac{d
\phi^i}{d\sigma}$. This choice ensures that the bosonic
action will be quadratic in derivatives which is a minimum
requirement if this model is ultimately to be interpreted as
a supersymmetric theory. In this case the action reads
\begin{equation}
S=\alpha\int d\sigma
\left(\frac{1}{2}g_{ij}\frac{d\phi^i}{d\sigma}\frac{d\phi^j}{d\sigma}
-\eta_i \frac{D}{D\sigma} \psi^i+
\frac{1}{4}R_{jlmk}\eta^j\eta^l\psi^m\psi^k\right)
\end{equation}
where
\begin{equation}
\frac{D}{D\sigma}\psi^i=\frac{d}{d\sigma}\psi^i+\Gamma^i_{kj}\frac{d\phi^k}{d\sigma}\psi^j
\end{equation}
is the pullback of the target space covariant derivative. This action 
corresponds to supersymmetric quantum mechanics (in Euclidean time) in which
the fields take their values on some non-trivial target space. It is
a straightforward generalization of the model considered earlier \cite{qm}.
It is clear that the special choice of gauge condition
we made in making contact with the supersymmetric theory corresponds to
finding the Nicolai map for the supersymmetric theory \cite{nic}.

Perhaps the most important feature of this
continuum construction of the (twisted) supersymmetric
model is that it can be transcribed trivially to the
lattice {\it without mutilating the fermionic symmetry} -- simply by replacing
the continuum derivative in $N^i=\frac{d\phi^i}{d\sigma}$
by a suitable finite difference operator defined on
a one dimensional lattice in which the continuum
coordinate $\sigma$ is replaced by the discrete index $t$ where
$\sigma=ta$ with
$t=1\ldots N$ and $a$ is the lattice
spacing. 
\begin{equation}
N^i\to \frac{1}{a}\Delta_{tt^\prime}^+\phi^i_{t^\prime}=\frac{1}{a}\left(
\delta_{t^\prime,t+a}-\delta_{t^\prime,t}\right)\phi^i_{t^\prime}
\end{equation}
In this expression we employ a {\it forward} lattice difference operator
to ensure no doubles appear in the spectrum of bosonic states. Notice that
the (twisted) supersymmetry then guarantees that no such states can 
appear in the fermion spectrum either. This statement is certainly
true classically, however we believe that it is also remains true when
quantum fluctuations are taken into account.
Replacing continuum integrals with
sums over lattice points $x$ in the usual manner we
are led to a lattice action of the form.
\begin{equation}
S=\alpha\sum_x\left(\frac{1}{2}g_{ij}\Delta^+\phi^i\Delta^+\phi^j
-\eta_i D \psi^i+
\frac{1}{4}R_{jlmk}\eta^j\eta^l\psi^m\psi^k\right)
\end{equation}
where
\begin{equation}
D\psi^i=\Delta^+\psi^i+\Gamma^i_{jk}\left(\Delta^+\phi^j\right)\psi^k
\end{equation} 
While this
lattice action is invariant under the twisted supersymmetry
it is no longer generally coordinate invariant in
the target space. This derives from the fact that
the lattice expression for $N^{i}\left(\phi\right)$ no longer
yields a target space vector when the base space derivative is replaced
by a finite difference operator. However, the deviation of $N^{i}$ from a 
true vector field decreases smoothly to zero
as $\alpha\to\infty$ and the
continuum limit is approached. Thus in the vicinity of such a continuum
fixed point it appears that the lattice theory approaches a generally
covariant continuum theory possibly perturbed by irrelevant operators. We are 
currently studying
this issue in more detail. 
 
Notice that the identification of this model with supersymmetric
quantum mechanics is very straightforward in the case of a one dimensional 
base space --
the twisting procedure is trivial in this case. It merely requires
one to identify the ghost and antighost with the physical ``fermion''
fields $\Psi$, $\overline{\Psi}$ respectively.
In higher dimensions the identification is somewhat more complicated
since the ghost and antighost will turn out to be related
to different chiral components of the physical fermions. 
Notice also that to make contact with the supersymmetric
theory we {\it do not} impose the usual physical state conditions
associated with BRST quantization - namely that
the BRST charge annihilate the physical states. Thus
the model is {\it not} topological. Nevertheless, the BRST symmetry,
is exact and will guarantee the vanishing of a set of
associated Ward identities corresponding to the twisted supersymmetry.
We now turn to the generalization of this model to describe a field
theory in two dimensions.

\section{Twisted two dimensional sigma models}

The model we discussed in the last section is sufficiently general that
it can be used to generate theories defined on a two
dimensional base space. In essence all we have to do is to
replace the ``gauge condition'' $N^i(\phi)$ by some suitable
function. To lead to a quadratic bosonic action it should involve
a single derivative of the field $\phi$ with respect to the base 
space coordinates. A simple guess for the continuum form would be
\begin{equation}
N^{i\alpha}=\partial^\alpha\phi^i
\end{equation}
Here, the vector $N^i$ must pick up an extra index $\alpha=1,2$
corresponding to the
two possible directions in the base space. Notice that the presence of
this extra index implies that both antighost and multiplier
field also pick up an extra index $\eta_i\to \eta_{i\alpha}$ and
$B_i\to B_{i\alpha}$. It is easy to verify that the fermionic transformations
are unaffected by the addition of this
extra index as is the nilpotent property of the BRST operator
corresponding to those transformations. Actually this choice will not
do. It is clear that if we are to arrive at a twisted supersymmetric
model the number of degrees of freedom carried by the antighost must
match that of the ghost field (in the end they will turn out to correspond to 
different chiral components of the physical fermions).
Thus we must require the antighost
$\eta$ and
multiplier field $B$ to satisfy some condition which halves their
number of degrees of freedom. The natural way to do this is to introduce
projection operators $P^{\left(-\right)}$ and $P^{\left(+\right)}$ and
require that $\eta^{i\alpha}$ and $B^{i\alpha}$ satisfy 
certain self-duality
conditions
\begin{eqnarray}
P^{\left(-\right)}\eta&=&0\nonumber\\
P^{\left(+\right)}\eta&=&\eta
\label{self-dual}
\end{eqnarray}
One choice for these projectors is
\begin{equation}
{P^{i\alpha}_{j\beta}}^{\left(\pm\right)}=
\frac{1}{2}\left(\delta^i_j\delta^\alpha_\beta\pm
J^i_j\epsilon^\alpha_\beta\right)
\end{equation}
Here, $J^i_j$ must be a globally defined tensor field
on the
target space and $\epsilon^\alpha_\beta$ is the usual antisymmetric
matrix with constant coefficients
(at this point we shall think of the base space as
flat).  
In order that 
$P^{\left(+\right)2}=P^{\left(+\right)}$,
$P^{\left(-\right)2}=P^{\left(-\right)}$ and
$P^{\left(+\right)}P^{\left(-\right)}=P^{\left(-\right)}P^{\left(+\right)}=0$ 
the tensor $J^i_j$ must be antisymmetric and square to minus the identity. 
Manifolds
possessing such a structure are 
called {\it almost complex} and have
even dimension. At this point we must be careful to make sure that
the BRST transformations we introduced earlier eqn.~\ref{BRST} 
are compatible with
these self-duality conditions given in eqn.~\ref{self-dual}.
Explicitly, the BRST transformations will need no modification if
\begin{equation}
\delta\left({P^{i\alpha}_{j\beta}}^+\eta^{j\beta}\right)=\delta\eta^{i\alpha}
\end{equation}
This latter condition is satisfied provided the almost complex structure is
covariantly constant $\nabla_kJ^i_j=0$. Manifolds possessing
a covariantly constant almost complex structure are termed 
K\"{a}hler.
Using these projectors the continuum gauge condition now becomes
\begin{equation}
N^{i\alpha}={P^{i\alpha}_{j\beta}}^+ \partial^\beta\phi^j
\label{gaugefix2d}
\end{equation}
Just as in the previous section we choose a 
a pure gauge-fixing term as action 
\begin{equation}
S=\alpha\delta \int d^2\sigma \sqrt{h} \left(\eta_{i\alpha}\left(N^{i\alpha}-\frac{1}{4}B^{i\alpha}\right)\right)
\label{2dgaugefermion}
\end{equation}
where the change in the coefficient of the multiplier term is simply to
generate a conventional normalization for the boson kinetic term. In this
expression we have 
now allowed for a general metric $h_{\alpha\beta}$ in the base space
(in which case we must interpret $\epsilon^\alpha_\beta$ as an almost
complex structure on the base space).
Carrying out the variation and integrating out the multiplier field leads
to the on shell action
\begin{eqnarray}
S&=&\alpha\int d^2\sigma \sqrt{h}\left(
\frac{1}{2}h^{\alpha\beta}g_{ij}\partial_{\alpha}\phi^i\partial_{\beta}\phi^j+
\frac{1}{2}\epsilon^{\alpha\beta}J_{ij}\partial_{\alpha}\phi^i\partial_{\beta}\phi^j\right.\nonumber\\
&-&\left.\eta_{i\alpha}\nabla_{k}
N^{i\alpha}\psi^k+\frac{1}{8}R_{ijkm}\eta^{i\alpha}\eta^{j}_{\alpha}\psi^k\psi^m\right)
\end{eqnarray}
In this expression $g_{ij}$ denotes, as before, the metric on the target space.
Notice that this invariance of the action under
the BRST symmetry is true independent of the choice of
either base or target space metric provided they are both K\"{a}hler. This
observation lies at the heart of the use of these models to
construct topological field theories. 
Furthermore, for such manifolds, the projector, buried in
the ghost-antighost term, may be 
taken out of the covariant derivative 
to act harmlessly on the
self-dual antighost on the left.
This term then takes the form
\begin{equation}
\eta_{i\alpha}D^\alpha\psi^i
\end{equation}
where the base space covariant derivative acting on the ghost field
arises from the pullback of the target space covariant derivative acting
on the gauge function. In detail
\begin{equation}
D^\alpha\psi^i=\partial^\alpha\psi^i+\Gamma^i_{jk}\partial^\alpha\phi^j\psi^k
\end{equation}
Furthermore, in the case of a K\"{a}hler manifold the
piece of the boson action involving $J^i_j$ is a topological
invariant. 
The physical interpretation of this model is clearer if we go to
complex or K\"{a}hler coordinates in the
target space. The original target space fields $\phi^i, i=1\ldots 2N$
are replaced with complex fields $\phi^I, I=1\ldots N$ and their complex conjugates
$\left(\phi^I\right)^*=\phi^{\overline{I}}$. In these coordinates the
only non-zero components of the metric take the form $g_{I\overline{J}}$
and $g_{\overline{I}J}=\left(g_{I\overline{J}}\right)^*$. Furthermore,
the tensor $J^i_j$ locally takes the form $i\delta^I_J$ for the
unbarred coordinates and $-i\delta^{\overline{I}}_{\overline{J}}$ for
the barred coordinates. The target space interval can be
written
\begin{equation}
ds^2=2g_{I\overline{J}}d\phi^Id\phi^{\overline{J}}
\end{equation}
Similarly we can adopt complex coordinates to parametrize the
base space which we can take
as flat. The original real coordinates $\sigma_1$, $\sigma_2$ are
replaced with the complex coordinates $\sigma_\pm=\sigma_1\pm i\sigma_2$
with $h_{+-}=h_{-+}=2$ and base space line element
$dl^2=2h_{+-}d\sigma^+d\sigma^-$
In these coordinates the gauge condition becomes simply the
Cauchy-Riemann condition
\begin{equation}
\partial_+\phi^I=0
\end{equation}
solutions of which are holomorphic functions.
In such complex coordinates the self-duality condition on the antighost field
$\eta$ implies
\begin{eqnarray}
\eta_-^I&=&0\nonumber\\
\eta_+^{\overline{I}}&=&0
\end{eqnarray}
Similarly the action takes the form
\begin{eqnarray}
S&=&\alpha\int d^2\sigma\left(
2h^{+-}g_{I\overline{J}}\partial_+\phi^I
\partial_-\phi^{\overline{J}}\right.\nonumber\\
&-&\left.h^{+-}g_{I\overline{J}}\eta^I_+D_-\psi^{\overline{J}}-
h^{+-}g_{\overline{I}J}\eta^{\overline{I}}_-D_+\psi^J+
\frac{1}{2}h^{+-}R_{I\overline{I} J\overline{J}}
\eta^I_+\eta^{\overline{I}}_-\psi^J\psi^{\overline{J}}\right)
\label{2daction}
\end{eqnarray}
Notice that the only non-zero terms of the Riemann tensor for K\"{a}hler
manifolds are precisely of the form shown and guarantee that this
term is real. The covariant derivatives acting on
the ghost fields are given by
\begin{equation}
D_+\psi^J=\partial_+\psi^J+\Gamma^J_{KL}\partial_+\phi^K\psi^L
\end{equation}
Again, for K\"{a}hler manifolds parametrized in complex coordinates 
the only non-zero
connection terms are of the form $\Gamma^I_{JK}$ and its
complex conjugate $\Gamma^{\overline{I}}_{\overline{J}\overline{K}}$.
In terms of these complex fields the BRST transformations become
\begin{eqnarray}
\delta\phi^I&=&\xi\psi^I\\
\delta\psi^I&=&0\\
\delta\eta^I_+&=&\xi\left(2\partial_+\phi^I+\Gamma^I_{JK}\eta_+^K\psi^J\right)
\label{2dbrst}
\end{eqnarray}
This action and associated
fermionic symmetry is in agreement with Witten's
original construction of topological sigma models restricted
to the case of K\"{a}hler target spaces
\cite{witten2}.

At this point the physical interpretation of the model is not yet
transparent. We need to rewrite the theory in terms of physical
fermion fields -- that is we need to {\it untwist} the theory.
The first stage in this process consists of recognizing that the 
the operator $\partial_+$ can be
thought of as the Weyl operator in Euclidean space. To see this consider
the chiral representation of the Dirac matrices in Euclidean space
\[\begin{array}{cc}
\gamma_1=\left(\begin{array}{cc}
0&1\\
1&0\end{array}\right)
&
\gamma_2=\left(\begin{array}{cc}
0&i\\
-i&0\end{array}\right) 
\end{array}
\]
Thus the free Dirac operator in this basis is just
\[\left(\begin{array}{cc}
0&\partial_+\\
\partial_-&0
\end{array}\right)\]
Since Weyl spinors are just complex numbers in two
dimensions we see that solutions of the gauge condition are in one-to-one
correspondence with solutions of the
Weyl equation for right handed spinors.
Indeed we see that the part of the action \ref{2daction} quadratic
in the ghost-antighost fields can be rewritten in this chiral
basis as
\begin{equation}
\overline{\Psi}^{I} -i\gamma \cdot D\Psi^{\overline{J}}g_{I\overline{J}}
\end{equation}
where 
\begin{equation}
\begin{array}{ccc}
\Psi^{\overline{J}}=\left(\begin{array}{c}\psi^{\overline{J}}\\
\frac{1}{2}i\eta^{\overline{J}}_-\end{array}\right)&\;\;\;&
\overline{\Psi}^{I}=\left(\begin{array}{c}\psi^I\\
\frac{1}{2}i\eta_+^I\end{array}\right)
\end{array}
\label{dirac}
\end{equation}
With this relabeling we can see that
the model may be identified with the usual
supersymmetric sigma model with $N=2$ supersymmetry with
the gauge parameter $\alpha$ reinterpreted as the physical
coupling $\alpha=\frac{4}{g^2}$ (see \cite{rev},\cite{shifman}).
The crucial
ingredient which made this final identification possible was the
requirement that the target manifold be K\"{a}hler. This allowed
the construction of a covariantly constant projection operator 
which was compatible with the BRST symmetry and which could be used to
define a self-dual gauge fixing condition
eqn.~\ref{gaugefix2d}. The structure of the latter then gave rise to
an equation for the ghost fields that could be interpreted as a
Dirac equation for a chiral fermion. 
Turning this argument around it is clear that {\it only} 
supersymmetric models with
K\"{a}hler target spaces can be {\it twisted} to reveal actions invariant
under a BRST symmetry.

It is clear that the twisting procedure for these
two dimensional sigma models can be
regarded as essentially a decomposition of the original Dirac field
into chiral components. To clarify this issue consider the
$N=2$ supersymmetry algebra
\[
\{Q_{\alpha +}, Q_{\beta -}\}=\gamma^\mu_{\alpha\beta}P_\mu
\]
where $+/-$ denote quantum numbers $+1/2$ and $-1/2$ of the internal
R-symmetry with generator $R$. 
The original rotation group has a single generator $J$ with
quantum numbers in a spinor representation $\alpha=+$ and $\alpha=-$. The
supercharges are then denoted $Q_{\pm\pm}$. When the theory is
twisted we replace the original generator of rotations $J$ by $J+R$.
It is now clear that two supercharges have spin zero
$Q_L=Q_{+-}$ and $Q_R=Q_{-+}$. Furthermore,
it follows from the original superalgebra that
these new charges obey the BRST-like algebra
$$Q_L^2=Q_R^2=\{Q_L,Q_R\}=0$$
We can then use say $Q_L$ to construct our twisted supersymmetric
action. The twisting procedure does not always
yield such a trivial change of variables. Consider, for example, four
dimensional super Yang Mills theories in which the fields
transform as representations of $SU(2)_L\times SU(2)_R\times SU(2)_I$
where $I$ labels the isospin or R-symmetry.
The twisting procedure amounts to constructing a new rotation group
$SU(2)_L\times SU(2)_I$ and decomposing all fields into
representations of this group. After this decomposition a scalar
nilpotent charge is produced corresponding to the trace of $Q^i_\alpha$
where $i$ and $\alpha$ are the original isospin and spinor indices.
In this case the twisted fermion fields correspond to fields which 
transform as a scalar, vector and selfdual tensor under the new
rotation group. We refer the reader to \cite{witten1} for
more details.

Returning now to the sigma models, it is trivial
to see that the invariance of the action under
the twisted supersymmetry
is retained when I replace the continuum action with an
appropriate lattice action.
However, unlike the case of quantum mechanics, we cannot do this
merely by replacing all continuum derivatives by forward difference
operators -- it is easy to see that the kernel of the (free) 2D
lattice Dirac operator constructed this way still contains extra states
which have no continuum interpretation\footnote{We thank
Joel Giedt and Erich Poppitz for this observation}. Instead we must proceed in
a way similar to the complex Wess-Zumino model \cite{wz2} and introduce
an explicit Wilson mass term. Of course it is not
obvious that the addition of such a term is compatible with
the topological symmetry in the case of a curved target space.
However, in \cite{pot, 2dtop} it was shown that indeed it is
possible to add potential terms to the twisted sigma models while
maintaining the topological symmetry.
In complex coordinates the possible terms are
\begin{equation}
\Delta S= \lambda^2
V^IV_I+\lambda^2\psi^I\nabla_IV_{\overline{J}}\psi^{\overline{J}}-
\frac{1}{4}h^{+-}\eta_+^I\nabla_IV_{\overline{J}}\eta_-^{\overline{J}}
\label{killing}
\end{equation}
Here, $V^I$ is a holomorphic Killing vector and $\lambda^2$ an
arbitrary parameter. A Wilson term would correspond to
the choice $V^I=im_W\phi^I$ where
\beq
m_W=\frac{1}{2}\left(\Delta^+_z\Delta^-_{\overline{z}}+
                     \Delta^+_{\overline{z}}\Delta^-_z\right)
\eeq
Many K\"{a}hler manifolds 
possess such a holomorphic Killing vector (for example the
$CP^N$ models considered in the next section). Thus for these models
the doubled modes can be removed without spoiling the $Q$-exactness of
the lattice action. 

In detail, the transcription for latticization now entails replacing the
continuum
derivative in the gauge condition by a {\it symmetric} 
finite difference operator 
\begin{equation}
N^I_+=\Delta^S_+\phi^I
\end{equation}
where $\Delta^S_+=\Delta^S_1+i\Delta^S_2$
and $\Delta^S=\frac{1}{2}\left(\Delta^+ +\Delta^-\right)$ together
with the appropriate potential terms following from eqn.~\ref{killing}
with $V^I=im_W\phi^I$.

The potential terms
should eliminate the doubles from both fermion and boson
sectors while
preserving lattice rotational invariance of the
theory at {\it any} lattice spacing. 
Thus far we have left the constant $\lambda^2$ free. In the topological
field theory language it corresponds to a gauge parameter. However,
if we untwist the model to a theory of Dirac fermions there is
a natural choice which preserves the Lorentz invariance of the model
$\lambda^2=h^{+-}=\frac{1}{2}$ for a flat base space.
Using this and the explicit form of the Killing vector allows
us to rewrite the additional piece in the action as
\beq
\frac{2}{\alpha}\Delta S=\left(m_W\phi^I\right)\left(m_W\phi_I\right)+
\overline{\Psi}_Iim_W\Psi^I
\eeq
Where the decomposition of the Dirac field in terms of the ghost and
antighost was given in eqn.~\ref{dirac}.

Finally, we turn to a discussion of the renormalization properties of
these two dimensional models. Specifically we would like to know whether
this lattice model approaches the continuum theory with the full $N=2$
supersymmetry automatically as the lattice spacing $a\to 0$.  
The requirement that the action is invariant under the twisted
supersymmetry ensures that any local counterterm induced via quantum effects
must take the form of a gauge fixing term\footnote{Actually, for
the model with potential terms $Q^2$ corresponds to a target space coordinate
transformation along the Killing vector. This is {\it not} an exact
invariance of the lattice action but yields a supersymmetry breaking
term which vanishes like $a^3$. In a superrenormalizable theory this
term cannot generate {\it relevant} supersymmetry breaking terms
and the above analysis is still essentially correct}. 
Furthermore, simple
power counting arguments lead us to conclude that the only (marginally)
{\it relevant} terms take the form
\begin{equation}
O=\delta\left(\eta^{i\alpha}f_{ij}\left(\phi\right)
\left(\partial_\alpha\phi^j+B^j\right)\right)
\end{equation}
where we have reverted to the original formulation involving real
fields. Notice that we have written this operator in continuum language
assuming a restoration of Poincar\'{e} invariance in the base space.
We have also assumed that the latter symmetry additionally enforces general
covariance in the target space as we discussed in the one dimensional case.
General covariance ensures that $f_{ij}$  
is a tensor which may then be taken to represent a quantum
renormalization of the target space metric tensor. This
counterterm structure would be consistent with
a lattice model which exhibits $N=1$ supersymmetry in the
continuum limit. The restoration of
full $N=2$ supersymmetry appears to require additional constraints. Luckily,
such constraints are present in the form of additional discrete
symmetries
of the lattice action. Consider the classical action
in K\"{a}hler form given in eqn.~\ref{2daction}. 
It is trivial to see that this action is also invariant under the transformations
\begin{eqnarray}
\psi^I&\to & i\psi^I\nonumber\\
\psi^{\overline{I}}&\to & -i\psi^{\overline{I}}\nonumber\\
\eta^I_+&\to & i\eta^I_+\nonumber\\
\eta^{\overline{I}}_-&\to & -i\eta^{\overline{I}}_-
\label{sym}
\end{eqnarray}
Actually, this additional symmetry arises from the K\"{a}hler structure
of the target space appearing in the classical action. In \cite{alvarez}
it was shown that the $N=2$ supersymmetric action was invariant
under the transformations $\Psi^i\to J^i_j\Psi^j$ where $\Psi^i$ is
the Dirac spinor introduced earlier in connection to the untwisting of
the model. The tensor $J^i_j$ is just the (covariantly constant) complex
structure characteristic of a K\"{a}hler manifold. After twisting and
adopting complex coordinates we obtain the
symmetry shown in eqn.~\ref{sym}. This additional symmetry of the lattice model then ensures that only
counterterms compatible with a K\"{a}hler target space survive in the
quantum effective action.
But as was shown in \cite{alvarez} any
model with $N=1$ supersymmetry and a K\"{a}hler target space automatically
possesses $N=2$ supersymmetry. Thus we expect that no additional
fine tuning is needed to regain the full supersymmetry of the continuum
model.

In the above argument we have only considered radiatively
induced operators
which take the form $\delta O$. One might worry that other invariant
operators which are not of this form could be produced
via quantum effects. In fact the list of such operators for the 2d
$\sigma$ model is very short -- it consists of just one element (see
\cite{witten2})
$$O=iJ_{ij}\psi^i\psi^j$$
In the continuum limit (where the Wilson term vanishes) the
chiral symmetry of the bare Lagrangian prohibits such a term and hence
we can safely neglect it in our analysis.
 
We turn now to an explicit example of these ideas -- the $O(3)$ supersymmetric
sigma model.

\section{Lattice $O(3)$ nonlinear sigma model}

The $CP^N$ models are perhaps the simplest example of
sigma models possessing K\"{a}hler target spaces. They have been
extensively studied in the literature \cite{alvarez,cpn,shifman}
in part because of their similarities to gauge models in
four dimensions.
These models contain $N$ complex scalar fields $u^I$ acting as
coordinates on a complex manifold whose metric is locally 
derived from the K\"{a}hler potential 
$
V=\ln{\left(1+u^Iu^{\overline{I}}\right)}
$
in the usual way
\begin{equation}
g_{I\overline{J}}=\frac{1}{2}\frac{\partial^2 }
{\partial u^I\partial u^{\overline{J}}}V\left(\phi^I,\phi^{\overline{I}}\right)
\end{equation}
The case $N=1$ is especially interesting as it corresponds to the
usual $O(3)$ supersymmetric sigma model \cite{shifman}. 
In this case the metric, connection and
curvature are easily verified to be
\begin{eqnarray}
g_{u\overline{u}}&=&\frac{1}{2\rho^2}\\
\Gamma^u_{uu}&=&g^{\overline{u}u}\partial_u g_{\overline{u}u}=-\frac{2\overline{u}}{\rho}\\
R_{\overline{u}u\overline{u}u}&=&g_{\overline{u}u}\partial_{\overline{u}}\Gamma^u_{uu}=-\frac{1}{\rho^4}
\end{eqnarray} 
where $\rho=1+u\overline{u}$.
In this case the supersymmetric lattice action including Wilson
terms takes the form
\begin{eqnarray}
S&=&\alpha\sum_{x}\left[\frac{1}{\rho^2}\Delta^S_+u\Delta^S_-\overline{u}+
\frac{1}{\rho^2}(m_Wu)(m_W\overline{u})
-\frac{1}{2\rho^2}\eta D^S_-\overline{\psi}
-\frac{1}{2\rho^2}\overline{\eta} D^S_+\psi\right.\nonumber\\
&+&\left.
\frac{1}{\rho^2}\psi im_W\overline{\psi}-\frac{1}{4\rho^2}\eta
im_W\overline{\eta}
-\frac{1}{2\rho^4}\overline{\eta}\eta\overline{\psi}\psi\right]
\end{eqnarray}
where a factor of two has been absorbed into the
coupling $\alpha$ and we have simplified our notation by replacing
$\eta_+\to\eta$ and $\overline{\eta}_-\to\overline{\eta}$. The explicit
form of the lattice covariant derivative is 
\begin{equation}
D^S_+=\Delta^S_+ -\frac{2\overline{u}}{\rho}\left(\Delta^S_+ u\right)
\end{equation}
To proceed further it is convenient to introduce an auxiliary field $\sigma$
to remove the quartic fermion term (this field $\sigma$ has nothing to do
with the original base space coordinates which have been replaced by
integer lattice coordinates $x$). Explicitly we employ the
identity
\begin{equation}
\alpha^N\int D\sigma e^{-\alpha\left(\frac{1}{2}\sigma\overline{\sigma}+
\frac{\sigma}{2\rho^2}\overline{\eta}\psi+
\frac{\overline{\sigma}}{2\rho^2}\eta\overline{\psi}\right)}=e^{\frac{\alpha}{2\rho^4}
\overline{\eta}\eta\overline{\psi}\psi}
\end{equation}
where $N$ is the number of lattice sites. Thus the partition function
of the lattice model can be cast in the
form
\begin{equation}
Z=\int DuD\sigma D\eta D\psi e^{-S\left(u,\sigma,\eta, \psi\right)}
\end{equation}
where the action is now given by
\begin{eqnarray}
S&=&\alpha\sum_{x}\left[\frac{1}{\rho^2}\Delta^S_+u\Delta^S_-\overline{u}+
\frac{1}{\rho^2}(m_Wu)(m_W\overline{u})+
\frac{1}{2}\sigma\overline{\sigma}\right.\nonumber\\
&-&\left.
\frac{1}{2\rho^2}\eta\hat{D}_-\overline{\psi}-
\frac{1}{2\rho^2}\overline{\eta}\hat{D}_+\psi+
\frac{1}{\rho^2}\psi im_W\overline{\psi}-\frac{1}{4\rho^2}\eta
im_W\overline{\eta}\right]
\label{o3act}
\end{eqnarray}
and the covariant derivative is modified to include a coupling to
the auxiliary field $\sigma$
\begin{equation}
\hat{D}_+=\Delta^S_+ -\frac{2\overline{u}}{\rho}\left(\Delta^S_+
u\right)+\sigma
\label{cov}
\end{equation}
Notice that the 
factor of $\alpha^N$ coming from the
introduction of the auxiliary field is canceled by a similar factor 
resulting from the
original gaussian integration over the multiplier field $B$. Thus, this
partition function should be {\it independent} of the coupling
constant $\alpha$ which plays the role of a gauge parameter from the
perspective of the BRST construction. We will present numerical results
confirming the $\alpha$ independence of $Z$ shortly.
Using the decomposition eqn.~\ref{dirac} the Dirac operator $M$ now takes the
form
\begin{equation}
M=\frac{1}{\rho^2}\left(\begin{array}{cc}
im_W&i\hat{D}_+\\
-(i\hat{D}_+)^\dagger&im_W
\end{array}\right)
\end{equation}
To simulate this model we have to reproduce the
fermion determinent arising after integrating out the
anticommuting fields. This leads to an effective action of the
form
\begin{equation}
S=\alpha S_B(u,\sigma)-\frac{1}{2}{\rm Tr}
\ln{\left(\alpha^2M^\dagger(u,\sigma)M(u,\sigma)\right)}
\end{equation}
where $S_B(u,\sigma)$ denotes the local bosonic pieces of the action and
we have shown the dependence on coupling $\alpha$ explicitly.
This representation of the fermion determinent requires that the latter
be positive. This is certainly true in the continuum limit when $M$ is
positive definite and we observe it to be true in all our simulations
at finite lattice spacing. 
For small lattices we have employed a HMC algorithm to yield an
exact simulation of this non-local effective action \cite{hmc}. This
algorithm requires a full inversion of the fermion matrix at every
time step and is prohibitively expensive for large lattices where
we have instead employed both a stochastic Langevin scheme and the so-called
R-algorithm to simulate the system \cite{lang}, \cite{ralg}. In these
cases physical quantities exhibit systematic errors $O(\Delta t)^2$. Typically,
for the data presented here we have used $\Delta=0.05$ which yields a
systematic error smaller than our statistical error.

A stringent test of the $\alpha$-independence of $Z$ is gotten by
measuring the expectation value of $S_B$. It should be clear that
\begin{equation}
-\frac{\partial\ln Z}{\partial\alpha}=0=\left<S_B\right>-\frac{2V}{\alpha}
\end{equation}
where $V$ denotes the number of lattice sites.
The results are given in the following table which shows data
for $\frac{\alpha}{2V}<S_B>$ from runs at a variety
of couplings $\alpha$ for lattices of size
$4\times 6$, and $8\times 8$. The runs comprise $1.7\times 10^4$ trajectories
and $1\times 10^4$ trajectories respectively. 
\TABULAR
{||l|l|l|l||}
{\hline
$\alpha$ &  $4\times 6$  &  $8\times 8$  \\\hline
$0.5$    &  $0.953(5)$    &  $-$         \\\hline
$1.0$    &  $0.969(4)$    &  $-$         \\\hline
$1.5$    &  $0.987(4)$    &  $-$         \\\hline
$2.0$    &  $0.999(4)$    &  $0.985(3)$  \\\hline
$2.5$    &  $1.002(4)$    &  $-$         \\\hline
$3.0$    &  $1.007(5)$    &  $0.995(3)$  \\\hline
}
{$\frac{\alpha<S_B>}{2V} $ vs $\alpha$}

Notice that while these numbers are statistically different from one
at small $\alpha$ they rapidly approach unity as $\alpha$ increases. This
is in agreement with our expectation that the Q-symmetry is only
broken by terms which vanish as a power of 
the lattice spacing. Most importantly no fine tuning is needed
to see a restoration of this symmetry in the continuum limit $\alpha\to\infty$. 
There is another way to understand this coupling constant
independence of
the partition function. In a conventional supersymmetric theory the partition
function gives a representation of the Witten index $W$. 
The latter can be written as
$$W={\rm Tr}\left(-1\right)^Fe^{-\alpha H}$$
where $F$ is the fermion number operator.
The topological
character of $W$ then relies on the bose-fermi pairing of all
states with non negative energy. Thus the $\alpha$ independence of $Z$,
guaranteed by the Q-exactness of the twisted action, appears to imply that
the Hamiltonian of the twisted system 
will indeed possess the exact degeneracy required
of a supersymmetric theory. 

As a further check on the supersymmetry of the model we have studied
the simplest two point functions involving boson and fermion fields.
The boson correlator $G^B(t)={\rm Re}\left(<u(t)\overline{u}(0)>\right)$ 
projected to zero spatial momentum is shown in figure 1. as a function
of the (Euclidean) time coordinate $t$ for
an $8\times 8$ lattice at $\alpha=2.0$. The curve shows the result
of a simple fit to the functional form
$a+b\cosh{\left(m^B\left(t-T/2\right)\right)}$ yielding $m^B=0.417(7)$ as an
estimate of the lowest lying boson mass state
($T$ denotes the length of the time axis). In this and the following
fits we exclude the $t=0$ data as it will contain contributions from
higher mass states.
The fermion correlator projected to zero spatial momentum  
$G^F_{ij}(t)=<\overline{\Psi}_i(t)\Psi_j(0)>$ 
should take the form
$$G^F_{ij}(t)=A(t-T/2)\delta_{ij}+iB(t-T/2)\epsilon_{ij}$$
where $A$ and $B$ are even and odd functions and we have taken the Dirac
gamma matrix in the time direction to correspond to $\sigma_2$. All
of our data very accurately reproduces this spinor structure.
Figure 2. shows a plot of ${\rm Im}G^F_{01}(t)$ as a function of $t$ again
for an $8\times 8$ lattice at $\alpha=2.0$. The solid curve shows a
simple fit to the form
$B=a\sinh{\left(m^F_{01}\left(t-T/2\right)\right)}$ with $m^F_{01}=0.41(1)$.
Similarly the function ${\rm Re}G^F_{00}(t)$ 
for the same run is shown in figure 3.
together with a simple fit to a hyperbolic cosine yielding 
a consistent estimate for the fermion mass $m^F_{00}=0.409(1)$. 
Thus the fits are consistent with the equality $m^B=m^F$ as required
by supersymmetry. 

\EPSFIGURE{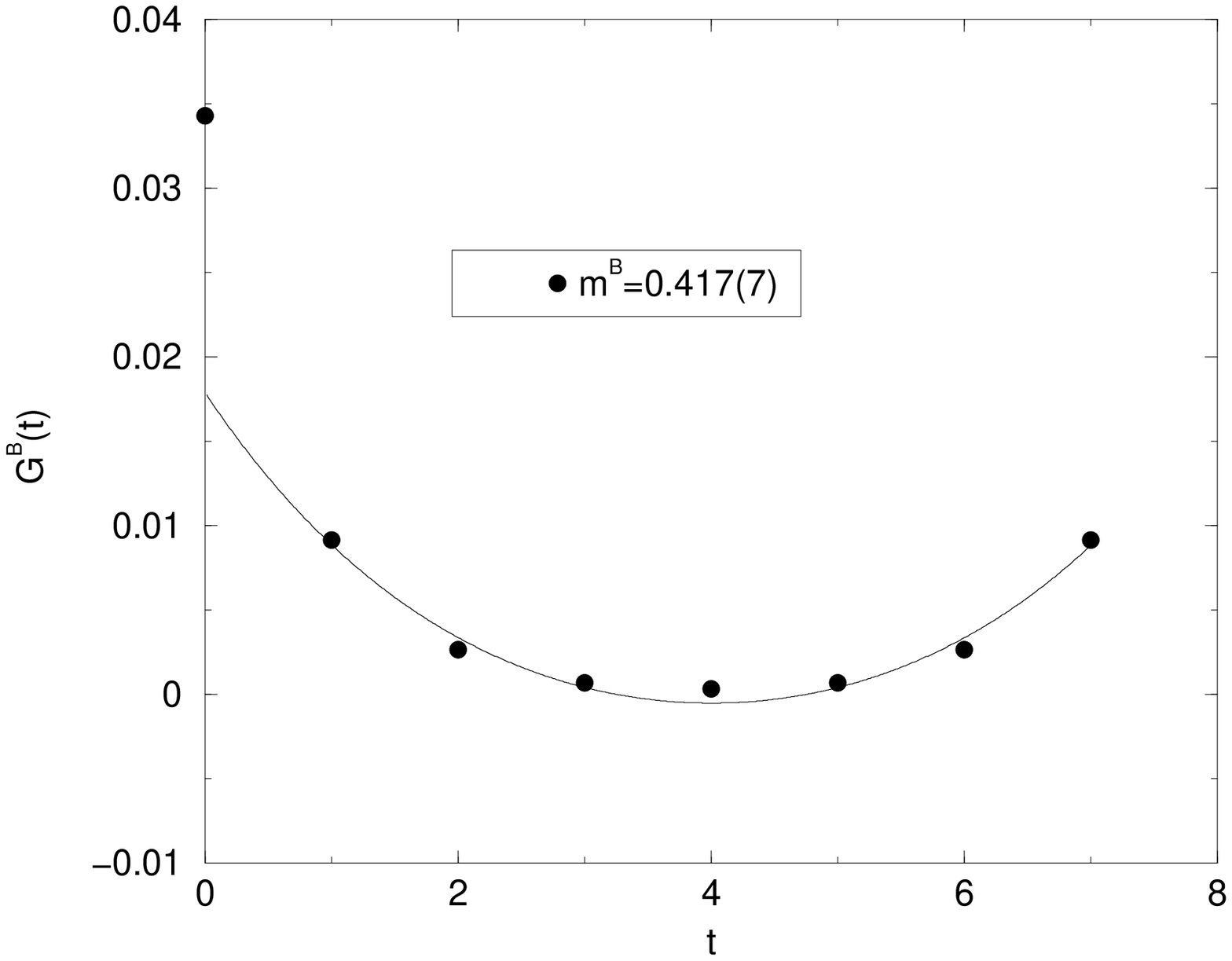,width=11cm}{$G^B(t)$ for $\alpha=2$, $L=8$}
\EPSFIGURE{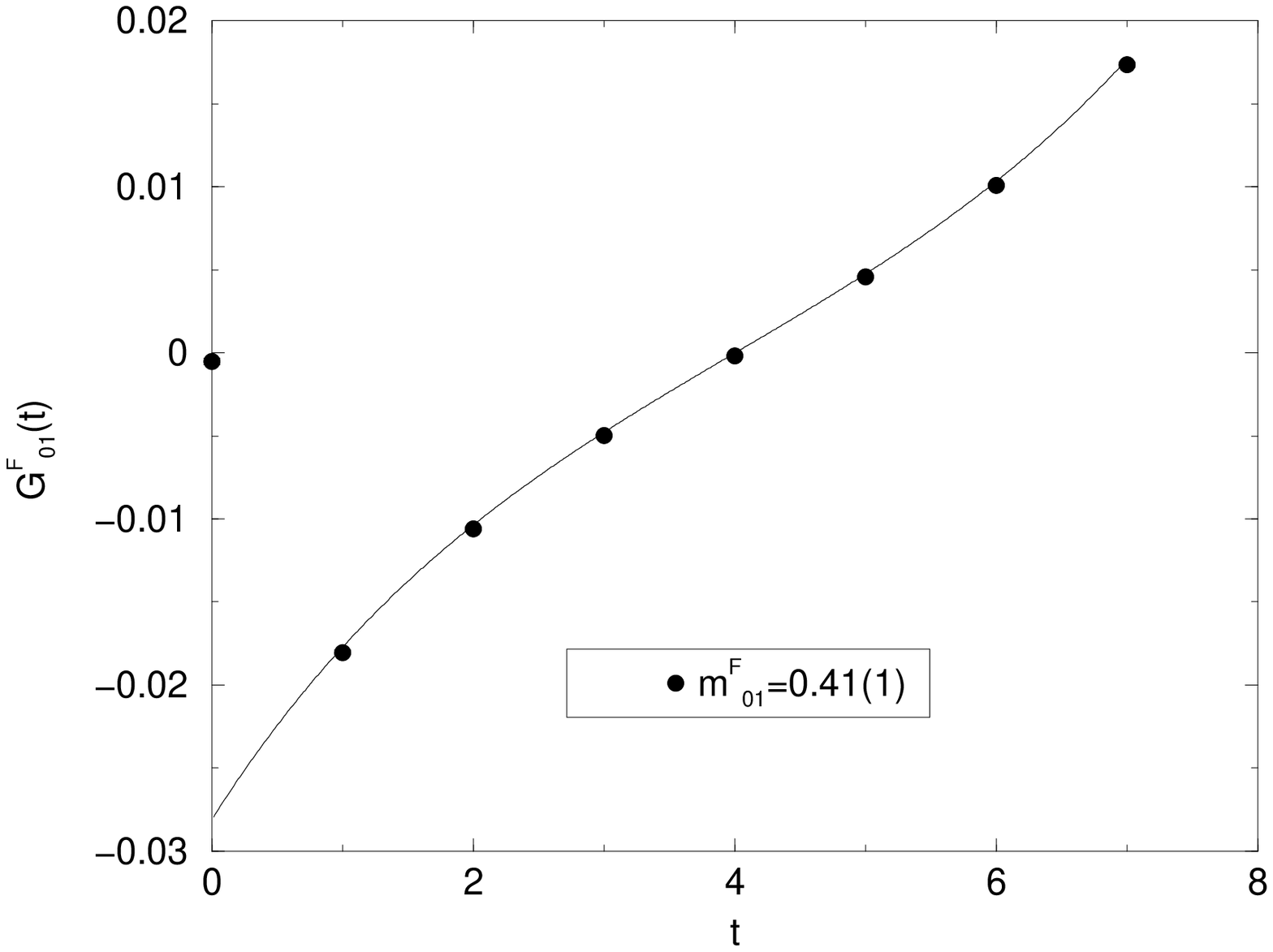,width=11cm}{$G^F_{01}(t)$ for $\alpha=2$, $L=8$}
\EPSFIGURE{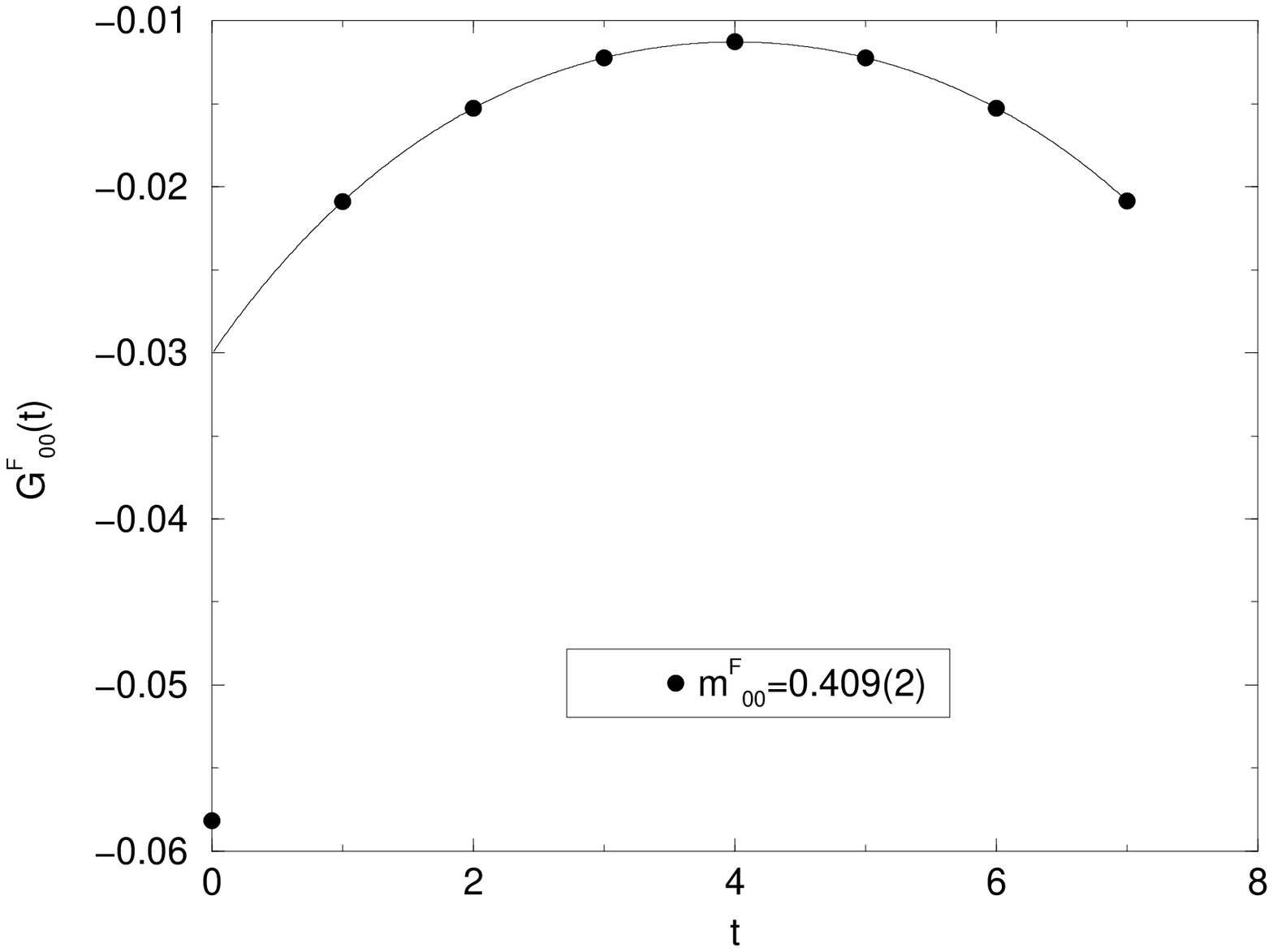, width=11cm}{$G^F_{00}(t)$ for $\alpha=2$, $L=8$}

We are currently extending these simulations to larger $\alpha$ and
bigger lattices to study the approach to the continuum limit. Notice that
a great deal is known theoretically about the mass spectrum of this
model in the continuum \cite{tim} and it would be very interesting to check 
the results of our simulations against this work.
Furthermore,
it is possible to write down lattice Ward identities for both the exact and
broken supersymmetries. Study of these should give information on the
question of the
restoration of the full $N=2$ supersymmetry in the limit of
vanishing lattice spacing. These questions are also under
investigation.

\section{Lattice Sigma Models in Four Dimensions}

Let us turn now to four dimensions. Again, the appropriate
gauge condition
$N^i\left(\phi\right)=0$ must take the form of a (base space) derivative
of the target space bosonic field $\phi^i$. To ensure that the
corresponding antighost $\eta_{i\alpha}$ and multiplier field $B_{i\alpha}$
possess
the correct numbers of degrees of freedom this gauge condition should
employ some projector to enforce a self-duality condition. In four
dimensions this condition should reduce the number of degrees of freedom
by a factor of four. In two dimensions the projector depended on
an {\it almost complex} structure for both base and target spaces both
of which were required to be K\"{a}hler.
The local
complex structure associated with a K\"{a}hler manifold then ensured that
the ghost-antighost action could be reinterpreted as action for
Dirac fermions in a chiral representation. 
In the
same way a local {\it almost quaternionic structure} is necessary to
allow for a representation of the four dimensional Dirac gamma matrix
algebra. Manifolds with an almost
quaternionic structure possess three independent tensors 
$\left(J^a\right)^i_j\;a=1\ldots 3$ satisfying the quaternionic algebra
\begin{equation}
J^aJ^b=-\delta^{ab}I+\epsilon^{abc}J^c
\label{quaternion}
\end{equation}
In terms of these tensors the projector we need looks like \cite{topsigma}
\begin{equation}
{P^{i\alpha}_{j\beta}}^+=\frac{1}{4}\left(\delta^i_j\delta^\alpha_\beta+
\left(J^a\right)^i_j \left(j^a\right)^\alpha_\beta \right)
\end{equation}
Here, the tensors $J^a$ are almost complex structures in the target
space and $j^a$ are the corresponding base space quantities.
The gauge condition in the continuum now reads
\begin{equation}
N^{i\alpha}={P^{i\alpha}_{j\beta}}^+\partial^\beta\phi^j
\label{4dgaugefix}
\end{equation}
and as before the antighost and multiplier fields are required to
be self-dual under the action of this projector
\begin{equation}
P^+\eta=\eta
\end{equation}
In a way similar to two dimensions the requirement that the
self-duality conditions are compatible with the BRST symmetry 
requires all three almost complex structures to be covariantly
constant. Four dimensional manifolds admitting three such covariantly
constant tensors are called hyperK\"{a}hler.

It is straightforward to find an action corresponding to fixing this
gauge condition
\begin{equation}
\xi S=\alpha \d \int d^4\sigma \sqrt{h}
\eta_{i\alpha}\left(N^{i\alpha}-\frac{1}{8}B^{i\alpha}\right)
\end{equation}
Carrying out the
variation and integrating out the multiplier field
(taking into account the self-duality condition) yields the following
on shell action 
\begin{eqnarray}
S&=&\alpha\int d^4\sigma\sqrt{h}\left(
\frac{1}{2}h^{\alpha\beta}g_{ij}\partial_\alpha\phi^i\partial_\beta\phi^j+
\frac{1}{2}\left(j^a\right)^{\alpha\beta}\left(J^a\right)_{ij}
\partial_\alpha\phi^i\partial_\beta\phi^j\right.\nonumber\\
&-&\left.h^{\alpha\beta}g_{ij}\eta^i_\alpha
D_\beta\psi^j+
\frac{1}{16}R_{ijkl}h^{\alpha\beta}\eta^i_\alpha\eta^j_\beta\psi^k\psi^l\right)
\end{eqnarray}
Notice again the presence of the
base space covariant derivative in the ghost-antighost term.
For hyperK\"{a}hler manifolds the term in the bosonic action containing the
tensors $J^a$ and $j^a$ is again a topological invariant.
It is possible to {\it untwist} this model to produce a Lorentz
invariant theory of
Dirac fermions in a way completely analogous to the two dimensional
case.
To make this explicit it is convenient to introduce a notation
that naturally emphasizes the quaternionic nature of
the hyperK\"{a}hler space.

First, group the original $4N$ target fields in sets of four 
\begin{equation}
\underbrace{\phi^1,\phi^2,\phi^3,\phi^4}_{\Phi^I,I=1},
\underbrace{\phi^5,\phi^6,\phi^7,\phi^8}_{\Phi^I,I=2},...,
\underbrace{\phi^{4N-3},\phi^{4N-2},\phi^{4N-1},\phi^{4N}}_{\Phi^I,I=N}  
\end{equation}
then the gauge condition condition, 
${P^{i\alpha}_{j\beta}}^+\partial^\beta\phi^j=0$
can be rewritten (locally) as
\begin{eqnarray}\label{cc}
\left\{ \begin{array}{ccc}
\partial_\mu \Phi^{I}_\mu & = & 0\\
\partial_\mu \Phi^{I}_\nu-\partial_\nu \Phi^{I}_\mu-\epsilon_{\mu\nu\alpha\beta}
\partial_\alpha \Phi^{I}_\beta & = & 0
\end{array} \right.
\end{eqnarray}
($\Phi^I_\mu$ really means $\phi^{4*(I-1)+\mu}$ with $I=1,...,N$ and $\mu=1,2,3,4$).
To see that this is true
let us choose the following representation for the $J^a$'s 
(likewise for the $j^a$'s)  
\begin{eqnarray}
\begin{array}{ccc}
J^1=I=\left( \begin{array}{cc} i\sigma_2 & 0\\ 0 & i\sigma_2 \end{array}\right) & 
J^2=J=\left( \begin{array}{cc} 0 & -\sigma_1 \\ \sigma_1 & 0 \end{array}\right) &
J^3=K=\left( \begin{array}{cc} 0 & -\sigma_3 \\ \sigma_3 & 0 \end{array}\right)
\end{array}
\end{eqnarray}
Equations (\ref{cc}) are 
equivalent to the Cauchy-Fueter equations that generalize to
four dimensions the Cauchy-Riemann equations of
two dimensions \cite{topsigma}. Finally let us
adopt quaternionic or hyper-K\"ahler coordinates by defining
\begin{equation}\begin{array}{lr}
\Phi^I=-I\Phi^I_1-J\Phi^I_2-K\Phi^I_3+\Phi^I_4\;{\rm and}\; \partial_+=I\partial_1+J\partial_2+K\partial_3+\partial_4
\end{array}
\end{equation}
then equations (\ref{cc}) for the gauge condition take the simple form,
\begin{equation}
\partial_+\Phi^I=0
\end{equation}
Solutions to this equation define so-called {\it triholomorphic} functions.
The action for the 4D model may then be 
written in a way analogous to the two dimensional case eqn.~\ref{2daction}
by allowing all fields to take values in the quaternions and recognizing that
a single quaternionic antighost (and its conjugate)
$\eta^M_+=\eta^M_4+I\eta^M_1+J\eta^M_2+K\eta^M_3$ survives the self-dual
projection.

The physical interpretation of this model necessitates untwisting. The
procedure mimics the 2D case. First,
it is easy to show that the solutions to the 4D gauge condition 
correspond to solutions of the Weyl 
equation for right handed spinors just as in the
two dimensional case \cite{topsigma}.
Indeed, if we take the chiral Euclidean representation of the four
dimensional Dirac matrices,
\begin{eqnarray}
\begin{array}{cc}
\gamma_i=\left( \begin{array}{cc} 0 & -i\sigma_i \\ i\sigma_i& 0 \end{array} \right) &
\gamma_4=\left( \begin{array}{cc} 0 & 1 \\ 1 & 0 \end{array} \right)      
\end{array}
\end{eqnarray}
the free Dirac operator again becomes simply
\begin{equation}
\gamma.\partial = \left(
\begin{array}{cc} 0 & \partial_+ \\ \partial_- & 0 \end{array} \right)
\end{equation}
(where $\partial_-=-I\partial_1-J\partial_2-K\partial_3+\partial_4$) and we
adopt a common representation of the 
quaternionic structure in terms of the Pauli matrices
$I=-i\sigma_1$, $J=-i\sigma_2$ and $K=-i\sigma_3$.
Then the action of this 
operator on a right handed Weyl spinor,
\begin{equation}
\Psi^I=\left(
\begin{array}{c} 0 \\ 0 \\ \Psi^I_1\\ \Psi_2^I \end{array} \right)
\end{equation}
with,
\begin{eqnarray}
\Psi^I_1 &=& \Phi^I_4+i\Phi^I_3 \nonumber \\
\Psi^I_2 &=& -\Phi^I_2+i\Phi^I_1
\end{eqnarray}
again just yields the set of equations (\ref{cc}).
Thus, the local quaternionic structure of
the hyperK\"{a}hler target space automatically allows us
to encode the Clifford algebra of the Dirac matrices and
ensures that the ghost and antighost field
correspond to the chiral components of a single Dirac fermion. Thus the 
untwisted model corresponds to the usual $N=2$ theory
based on hyper-multiplets \cite{wein}

As with the one and two dimensional models it is trivial to discretize this
theory while preserving the twisted supersymmetry by simply replacing
the derivative operator appearing the gauge condition eqn.~\ref{4dgaugefix}
by a symmetric difference operator. Such a choice will give rise to
both bosonic and fermionic doublers. To remove such states from
the spectrum we again need to add a Wilson term in the form
of a simple mass operator to the theory.  By analogy with the
two-dimensional case it would seem that this would only be possible
if the target space of the model admits a {\it triholomorphic} Killing
vector linear in the field. Several examples of such theories are
discussed in \cite{sigpot} for hyperK\"{a}hler manifolds. For such
manifolds we can add a Wilson type operator of the form $m_W\phi$ while
maintaining the Q-exactness of the lattice action.

Following the arguments given earlier the exact supersymmetry will help
protect the theory against radiative corrections. 
Indeed, as in two dimensions,
the 4D lattice model will possess three additional
discrete symmetries corresponding to acting on the spinors of the theory by
each of the three independent complex structures. 
In the continuum these additional
discrete symmetries generate additional supersymmetries.
Clearly the requirement that the lattice effective action respect
these additional symmetries will force the target manifold to remain
hyperK\"{a}hler under renormalization. This in turn should
ensure the full $N=2$ supersymmetry is achieved at large
correlation length without further fine tuning.
\vfill

\section{Conclusions}
In this paper we have shown how sigma models in one, two and
four dimensions with $N=2$ supersymmetry
may be discretized in a way which preserves an exact 
supersymmetry. 
The argument relies on the well known twisting procedure
used to produce topological field theories in the continuum.
Here, this argument is turned around and topological
constructions are used to generate twisted supersymmetric theories
which may then be discretized 
while preserving an exact (twisted) supersymmetry. The twisted
supersymmetry takes the form of a BRST symmetry and the
resulting twisted supersymmetric action resembles a gauge fixing term. 
Notice that while the method we use to construct our
lattice actions parallels
continuum topological field theory constructions our theories
are {\it not} topological. They are to be regarded as simply relabeled
forms of conventional supersymmetric actions. Indeed, we
have untwisted the models explicitly to reveal
the physical fermion interpretation. The ghost and anti-ghost fields
just correspond to different chiral components of a Dirac fermion
field.
Notice that from the point of
view of simulation this untwisting procedure
is somewhat unnecessary - the twisted fields generate precisely
the same fermion determinant as the physical fields and it is only
this object that we must replicate in any simulation.

The models we construct are local and, at least for
target spaces possessing suitable holomorphic Killing vectors,
can be rendered free of doubling
problems while preserving the Q-exactness of the lattice action.
Our preliminary numerical results for the $CP^1$ model
lend support to these claims.
Furthermore, it seems plausible (though we have no proof) that the
restoration of Poincar\'{e} invariance leads to a restoration
of target space reparametrization invariance in these theories. If this
is so then we have argued that
the exact supersymmetry together with certain discrete
symmetries should automatically lead to a restoration of full $N=2$
supersymmetry in the continuum limit without additional
fine tuning.
We are
currently investigating these issues in more detail.

Of course one of the most important questions is whether these ideas
may be extended to theories with a gauge symmetry.
In the continuum it is indeed possible to construct twisted versions
of a variety of super Yang Mills model (the original example being afforded
by $N=2$ theory in four dimensions \cite{witten1}). However, there is an
important distinction between these twisted gauge models and
both the sigma models described here and the Wess-Zumino models that were considered
earlier \cite{top}. The relevant BRST symmetry is now nilpotent only
up to gauge transformations -- so if we want to latticize such models
it appears we must be careful to preserve gauge invariance. While this
manuscript was being prepared we received a preprint \cite{sugino}
in which significant progress in this direction had been achieved.
\vfill
\pagebreak

\appendix 
\section{Appendix}

In this appendix we want to show that
\begin{eqnarray}
\d^2\eta_i &=& 0 \nonumber \\
\d^2 B_i &=& 0
\end{eqnarray}
indeed from (\ref{aa}),
\begin{eqnarray}
\d^2 \eta_i &=& \xi(\d B_i-\d\eta_j \Gamma^j_{ik}\psi^k-\eta_j \d\Gamma^j_{ik}\psi^k)\nonumber \\
&=& \xi \left\{ \chi(B_j\Gamma^j_{ik}\psi^k-\frac{1}{2}\eta_jR^j_{\;ilk}\psi^l\psi^k)-\chi(B_j-\eta_l
\Gamma^l_{jm}\psi^m)\Gamma^j_{ik}\psi^k-\eta_j\partial_m\Gamma^j_{ik}\chi\psi^m\psi^k \right\}\nonumber \\
&=& -\frac{1}{2}\xi\chi \eta_jR^j_{\;ilk}\psi^l\psi^k+\xi\chi\eta_j(\Gamma^j_{lm}\Gamma^l_{ik}
+\partial_m\Gamma^j_{ik})\psi^m\psi^k \nonumber
\end{eqnarray}
but,
\begin{eqnarray}
\label{bb}
(\Gamma^j_{lm}\Gamma^l_{ik}
+\partial_m\Gamma^j_{ik})\psi^m\psi^k &=& \frac{1}{2}(\partial_m\Gamma^j_{ik}-\partial_k\Gamma^j_{im}
+\Gamma^j_{lm}\Gamma^l_{ik}-\Gamma^j_{lk}\Gamma^l_{im})\psi^m\psi^k \nonumber\\
&=&\frac{1}{2}R^j_{\;imk}\psi^m\psi^k
\end{eqnarray}
where $\{\psi^m,\psi^k\}=0$ has been used. Thus,
\begin{eqnarray}
\d^2\eta_i &=& -\frac{1}{2}\xi\chi \eta_jR^j_{\;ilk}\psi^l\psi^k
+\frac{1}{2}\xi\chi \eta_jR^j_{\;ilk}\psi^l\psi^k \nonumber \\
&=& 0
\end{eqnarray}
now consider $\d^2B_i$,
\begin{eqnarray}
\d^2B_i &=& \xi \left\{ \d B_j \Gamma^j_{ik}\psi^k+B_j\d \Gamma^j_{ik} \psi^k-\frac{1}{2} 
\d \eta_j R^j_{\;ilk} \psi^l\psi^k-\frac{1}{2}\eta_j\d R^j_{\;ilk}\psi^l\psi^k\right\}\nonumber \\
&=& \xi \left\{ \c(B_l\Gamma^l_{jm}\psi^m-\frac{1}{2}\eta_nR^n_{\;jlm}\psi^l\psi^m)\Gamma^j_{ik}\psi^k
+ B_j\partial_m\Gamma^j_{ik}\c \psi^m \psi^k \right. \nonumber \\
&& \left.
-\frac{1}{2} \c (B_j-\eta_n\Gamma^n_{js}\psi^s) R^j_{\;ilk}\psi^l\psi^k 
-\frac{1}{2} \eta_j \partial_mR^j_{\;ilk}\c \psi^m\psi^l\psi^k \right\} \nonumber \\
&=& \xi\c B_j(\Gamma^j_{lm}\Gamma^l_{ik}+\partial_m\Gamma^j_{ik})\psi^m\psi^k
-\frac{1}{2}\xi\c B_jR^j_{\;ilk}\psi^l\psi^k \nonumber \\
&& +\frac{1}{2} \xi \c \eta_j(\partial_m R^j_{\;ilk}+R^j_{\;nlm}\Gamma^n_{ik}+\Gamma^j_{nm}R^n_{\;ilk})\psi^m
\psi^l \psi^k
\end{eqnarray}
using (\ref{bb}), the two first terms cancel each other. The last term can be written as,
\begin{eqnarray}
\eta_j(\partial_m R^j_{\;ilk}+R^j_{\;nlm}\Gamma^n_{ik}+\Gamma^j_{nm}R^n_{\;ilk})\psi^m
\psi^l \psi^k = \eta_j(\partial_m R^j_{\;ilk}-\Gamma^n_{im}R^j_{\;nlk}+\Gamma^j_{nm}R^n_{\;ilk})\psi^m
\psi^l \psi^k \nonumber
\end{eqnarray}
(where we used the anti-symmetry of the Grassman's),
\begin{eqnarray}
\hspace{1 in} = \eta_j(\partial_m R^j_{\;ilk}+\Gamma^j_{nm}R^n_{\;ilk}-\Gamma^n_{im}R^j_{\;nlk}
-\Gamma^n_{ml}R^j_{\;ink}-\Gamma^n_{mk}R^j_{\;iln})\psi^m \psi^l \psi^k
\end{eqnarray}
(as the two added terms vanish when contracted with $\psi^m\psi^l\psi^k$),
\begin{eqnarray}
& = &  \eta_j \nabla_mR^j_{ilk}\psi^m\psi^l\psi^k \nonumber \\
& = & \frac{1}{3} \eta_j(\nabla_mR^j_{ilk}+\nabla_lR^j_{ikm}+\nabla_kR^j_{iml})\psi^m\psi^l\psi^k \nonumber \\
& = & 0
\end{eqnarray}
(by Bianchi identity). Thus $\d^2 = 0$ i.e. is a nilpotent operation on the multiplet $(\phi,\eta,B,\psi)$.\\

\acknowledgments
The
authors would like to thank Joel Rozowsky for numerous
discussions during the early stages of this work. This work
was supported in part by DOE grant DE-FG02-85ER40237.


\begin{thebibliography}{99}
\bibitem{wein} S. Weinberg, The Quantum Theory of Fields III (Cambridge University Press, 2000) 
\bibitem{prop} N. Seiberg and E. Witten, Nucl. Phys. B431 (1994) 484.
\bibitem{M} J.M. Maldecena, Adv. Theor. Math. Phys. 2 (1998) 231 [Int. J. Theor.
Phys 38 (1999) 1113].
\bibitem{old} M Golterman and D. Petcher, Nucl. Phys. B319 (1989) 307.\\
S. Elitzur and A. Schwimmer, Nucl. Phys. B226 (1983) 109.\\
N. Sakai and M. Sakamoto, Nucl. Phys. B229 (1983) 173.\\
J. Bartels and J. Bronzan, Phys. Rev. D28 (1983) 818\\
T. Banks and P. Windey, Nucl. Phys. B198 (1983) 68
\bibitem{recent} Y. Kikukawa and Y. Nakayama, Phys. Rev. D66 (2002) 094508.\\
Kazuo Fujikawa, Phys. Rev. D66 (2002) 074510.\\
J. Nishimura, S. Rey and F. Sugino, JHEP 0302 (2003) 032\\
J. Giedt, E. Poppitz and M. Rozali, JHEP 0303 (2003) 035\\
J. Giedt, Nucl. Phys.B 668 (2003) 138\\
J. Giedt, The fermion determinant in (4,4) 2d
lattice super Yang Mills, hep-lat/0307024 \\
J. Nishimura, Phys. Lett. B406 (1997) 215\\
S. Catterall and S. Karamov, Phys. Rev. D 68 (2003) 014503\\
W. Bietenholtz, Mod. Phys. Lett. A14 (1999) 51.
\bibitem{feo} A. Feo, Supersymmetry on the Lattice, hep-lat/0210015\\
\bibitem{kaplan} D. B. Kaplan, Recent Developments in Lattice Supersymmetry,
hep-lat/0309099
\bibitem{ven} C. Curci and G. Veneziano, Nucl. Phys. B292 (1987) 555.
\bibitem{montway}I. Montvay, Nucl. Phys. B466 (1996) 259. \\
G.T. Fleming, J.B. Kogut and P.M. Vranas, Phys. Rev D 64 (2001) 034510.
\bibitem{qm}
S. Catterall and E. Gregory, Phys. Lett. B487 (2000) 349.
\bibitem{wz2}
S. Catterall and S. Karamov, Phys. Rev. D65 (2002) 094501.
\bibitem{kap1}
D.B. Kaplan, E. Katz and M. Unsal, JHEP 0305 (2003) 037.
\bibitem{kap2} A.G. Cohen, D.B. Kaplan, E. Katz, M. Unsal, 
JHEP 0308 (2003) 024   
\bibitem{kap3} A.G. Cohen, D.B. Kaplan, E. Katz, M. Unsal, hep-lat/0307012.
\bibitem{top} S. Catterall, JHEP 0305 (2003) 038. 
\bibitem{rev} D. Birmingham, M. Blau, M. Rakowski and G. Thompson, Phys. Reps.
209, nos. 4 and 5 (1991) 129-340.
\bibitem{rev2} J.M.F. Labastida and C. Lozano, hep-th/9709192.
\bibitem{witten1} E. Witten, Comm. Math. Phys. 117 (1988) 353.
\bibitem{singer} L. Baulieu and I. Singer, Comm. Math. Phys. 125 (1989) 227. 
\bibitem{nic} H. Nicolai, Phys. Lett. B89 (1980) 341.
\bibitem{witten2} E. Witten, Comm. Math. Phys. 118 (1988) 411 
\bibitem{topsigma} D. Anselmi, P. Fr\`e, hep-th/9306080
\bibitem{pot} J. M. F. Labastida and P. M. Llatas, Phys. Lett. B271 (1991) 101.
\bibitem{2dtop}J. M. F. Labastida and P. M. Llatas, Nucl. Phys. B379 (1992) 220.
\bibitem{alvarez} L. Alvarez-Gaume, D.Z. Freedman, Comm. Math. Phys. 80 (1981) 443
\bibitem{cpn} B. Zumino, Phys. Lett. 87B (1979) 203
\bibitem{shifman} V. A. Novikov, M. Shifman, A. Vainshtein, V. Zakharov,
Phys. Rep. 116: 103 (1984)
\bibitem{sigpot} L. Alvarez-Gaume and D. Z, Freedman, Comm. Math. Phys. 91 (1983).
\bibitem{hmc} S. Duane, A. Kennedy, B. Pendleton and D. Roweth, Phys. Lett.
B195B (1987) 216. 
\bibitem{lang} G. Batrouni, G. Katz, A. Kronfeld, G. Lepage, B. Svetitsky and
K.Wilson, Phys. Rev. D32 (1985) 2736.
\bibitem{ralg} S. Gottlieb, W. Liu, D. Toussaint, R. Renken, R. Sugar, Phys. Rev. D35 (1987) 2531. 
\bibitem{tim} T. Hollowood and J. Evans, Phys. Lett. B343 (1995) 198.
\bibitem{sugino} F. Sugino, hep-lat/0311021.

\end{thebibliography}
\end{document}